\documentclass[pre,aps,showpacs,12pt]{revtex4}
\usepackage{epsfig}
\begin{document}

\title{On the nature of ferromagnetism in diluted magnetic
semiconductors: GaAs:Mn, GaP:Mn.}
\author{V.A. Ivanov$^{1,2}$, P.M. Krstaji\'{c}$^{1}$, F.M. Peeters$^{1}$,
V. Fleurov$ ^{3}$, and K. Kikoin$^{4}$}
\date{\today}

\begin{abstract}
A microscopic Hamiltonian for interacting manganese impurities in diluted
magnetic semiconductors (DMS) is derived. It is shown that in \textit{p}%
-type III-V DMS the indirect exchange between Mn impurities has similarities
with the Zener mechanism in transition metal oxides. Here the mobile holes
and localized states near the top of the valence band play the role of
unoccupied oxygen orbitals which induce ferromagnetism. The Curie
temperature estimated from the proposed \textit{kinematic exchange} agrees
with recent experiments on GaAs:Mn. The model is also applicable to the
GaP:Mn system.
\end{abstract}

\affiliation{$^{1}$Department Natuurkunde , Universiteit Antwerpen
(UIA), Universitetsplein 1, B-2610 Antwerpen, Belgium \\
  $^{2}$ N.S. Korsakow Institute of General and Inorganic
Chemistry of the Russian Academy of Sciences, Leninski prospect
31, 117 907 Moscow, Russia\\
  $^{3}$School of Physics and Astronomy, Tel Aviv University, Ramat Aviv,
69 978 Tel Aviv, Israel\\
  $^{4}$Physics Department, Ben-Gurion University, 84 105 Beer-Sheva,
Israel}

\maketitle

The discovery of ferromagnetism (FM) with $T_{C}$=110K in Ga$_{0.947}$Mn$%
_{0.053}$As \cite{Ohno10} stimulated the systematic study of the III-V
dilute magnetic semiconductors (DMS). Recently, above room temperature FM
order was found in p-type GaP and GaN doped with Mn \cite{Reed}.

Most of the existing theories of FM in III-V DMS are based on
semi-phenomenological models, which postulate the existence of local
magnetic moments on the Mn sites, indirect exchange between these moments
and holes in the valence band of the host crystals (see, \textit{e.g.} \cite
{Dietl}) and sometimes emphasizes the role of shallow acceptor levels \cite
{Duga}. In this paper we present a \textit{microscopic model}, which takes
into account the origin of localized magnetic moments and shallow acceptor
levels induced by the Mn impurity and derive the effective \textit{kinematic}
\textit{exchange} from the generic two-impurity Hamiltonian. This mechanism
cannot be reduced to any previously proposed models, but it is based on
Zener's idea \cite{Zener} of double exchange via unoccupied $p$-orbitals.

In the present paper the basis for a microscopic description of an isolated
magnetic impurity is the Anderson Hamiltonian \cite{Anderson} which was
modified in Refs. \cite{fleurov1,Haldane} to the case of a semiconductor
host. A two-impurity generalization of the Anderson model for metals was
proposed in Ref. \cite{Alex} (hereafter referred as AA). An indirect
exchange between magnetic moments arises as a result of virtual electron
transitions into unoccupied states shared by two impurities.

It is known (see Refs. \cite{Zunger,Kikoin}) that $3d$ impurities create
both deep localized states in the energy gap and resonance states in the
valence and conduction bands. Three types of states are generated by the $3d$
impurities in zinc blende semiconductors \cite{Hemstreet,Picoli,Singh}:
non-bonding states retaining the angular $e$-symmetry of the states in a
cubic crystal field and two types of $t_{2}$ states (bonding and
antibonding). The latter arise due to a strong hybridization between the
atomic $t_{2}$ orbitals and the $p$-states of the same symmetry belonging
mainly to the heavy hole (\textit{hh}) band \cite{Zunger,Kikoin}. They are
called crystal field resonance (CFR, predominantly $d$-type) and dangling
bond hybrid (DBH, predominantly $p$-type). One of these states, as a rule,
gives an impurity level in the energy gap and another manifests itself as a
resonance within the valence band. According to its position in a series of
transition metal elements, the Mn-impurity in the Ga-site should have the
configuration $3d^{4}$. However, the Mn ion retains its fifth electron in
the $3d$ shell because of a specific stability of a high-spin half-filled
state $d^{5}$ and the impurity state is $\mathrm{Mn}(3d^{5}\bar{p})$, $\bar{p%
}$ is the bound hole state. Actually, in GaAs:Mn the mobile hole
concentration is about 30\% of the nominal Mn concentration \cite{omiya}.

The $pd$-hybridization together with the Anderson-Hubbard repulsion $U$ is
eventually the source of the magnetic interaction in DMS. To describe the
indirect exchange, we start with the single impurity resonance scattering
model \cite{fleurov1,Haldane} for $t_{2}$ electrons and generalize it to the
case of two impurities along the lines of the AA approach \cite{Alex}. The
'passive' non-bonding $e$-states contribute to the localized moment, but do
not participate in the indirect exchange. Therefore, the minimal
two-impurity Hamiltonian involves $t_{2}$-electrons:
\begin{equation}
H=\sum_{\mathbf{p},\sigma }\varepsilon _{\mathbf{p}}^{h}c_{\mathbf{p}h\sigma
}^{\dagger }c_{\mathbf{p}h\sigma }+\sum_{\mathbf{p},\sigma ;j}\left( V_{%
\mathbf{p}d}c_{\mathbf{p}h\sigma }^{\dagger }d_{i\sigma }e^{i\frac{\mathbf{p}%
}{\text{%
h\hskip-.2em\llap{\protect\rule[1.1ex]{.325em}{.1ex}}\hskip.2em%
}}\mathbf{\cdot R}_{j}}+h.c.\right) +\sum_{\sigma ,i}\left( E_{d}\hat{n}%
_{i}^{\sigma }+\frac{U}{2}\hat{n}_{i}^{\sigma }\hat{n}_{i}^{\bar{\sigma}%
}\right) ,  \label{And1}
\end{equation}
where only \textit{hh} states are retained in the band Hamiltonian. Here $c_{%
\mathbf{p}h\sigma }^{\dagger }(c_{\mathbf{p}h\sigma })$ is the creation
(annihilation) operator of a \textit{hh} with the momentum $\mathbf{p}$ and
spin $\sigma $. The second term describes the resonant impurity scattering
induced by the $pd$-hybridization $V_{\mathbf{p}d}$. The last term contains
the atomic energy levels of the localized $d$ electrons with the $t_{2}$%
-electron occupation operator $\hat{n}_{i}^{\sigma }=d_{i\sigma }^{\dagger
}d_{i\sigma }$ of the Mn impurity in the Ga sites labelled by $i=1,2$.

The system of Dyson equations for the d-electron Green's functions $%
G_{dii^{\prime }}^{\sigma }$ ($i,i^{\prime }=1,2$) has the following form:
\begin{equation}
G_{dii^{\prime }}^{\sigma }(\varepsilon )=g_{i\sigma }(\varepsilon )\left(
\delta _{ii^{\prime }}+V^{2}\sum_{j}L_{ij}(\varepsilon )G_{dji^{\prime
}}^{\sigma }(\varepsilon )\right) ,  \label{AAA1}
\end{equation}
where $g_{i\sigma }(\varepsilon )=(i\varepsilon -E_{d}-Un_{i}^{-\sigma
})^{-1}$ is the single site Green's function for a $t_{2}$ electron, $%
L_{ij}(\varepsilon )=\sum_{\mathbf{p}}e^{-i\frac{\mathbf{p}}{\text{%
h\hskip-.2em\llap{\protect\rule[1.1ex]{.325em}{.1ex}}\hskip.2em%
}}\mathbf{\cdot }(\mathbf{R}_{i}-\mathbf{R_{j}})}(i\varepsilon -\varepsilon
_{\mathbf{p}}^{h})^{-1}$ is the lattice Green's function for \textit{hh}.
The momentum dependence of the hybridization matrix elements is neglected,
\textit{i.e.} $V_{\mathbf{p}d}\approx V$. The solution of the system of Eqs (%
\ref{AAA1}) is $G_{dii}^{\sigma }=\left[ g_{j\sigma }^{-1}(\varepsilon
)-V^{2}L_{jj}^{\sigma }(\varepsilon )\right] /\mathsf{R}^{\sigma
}(\varepsilon ),~~G_{dij}^{\sigma }=V^{2}L_{ij}^{\sigma }(\varepsilon )/%
\mathsf{R}^{\sigma }(\varepsilon ),$ $(i=1,2;j=2,1)$ and the two-impurity
levels are found from
\begin{equation}
\mathsf{R}^{\sigma }(\varepsilon )=\prod_{i=1,2}\left[ g_{i\sigma
}^{-1}(\varepsilon )-V^{2}L_{ii}^{\sigma }(\varepsilon )\right]
-V^{4}L_{12}^{\sigma }(\varepsilon )L_{21}^{\sigma }(\varepsilon )=0.
\label{AAA2}
\end{equation}

\begin{figure}[h]
\includegraphics[width=14cm]{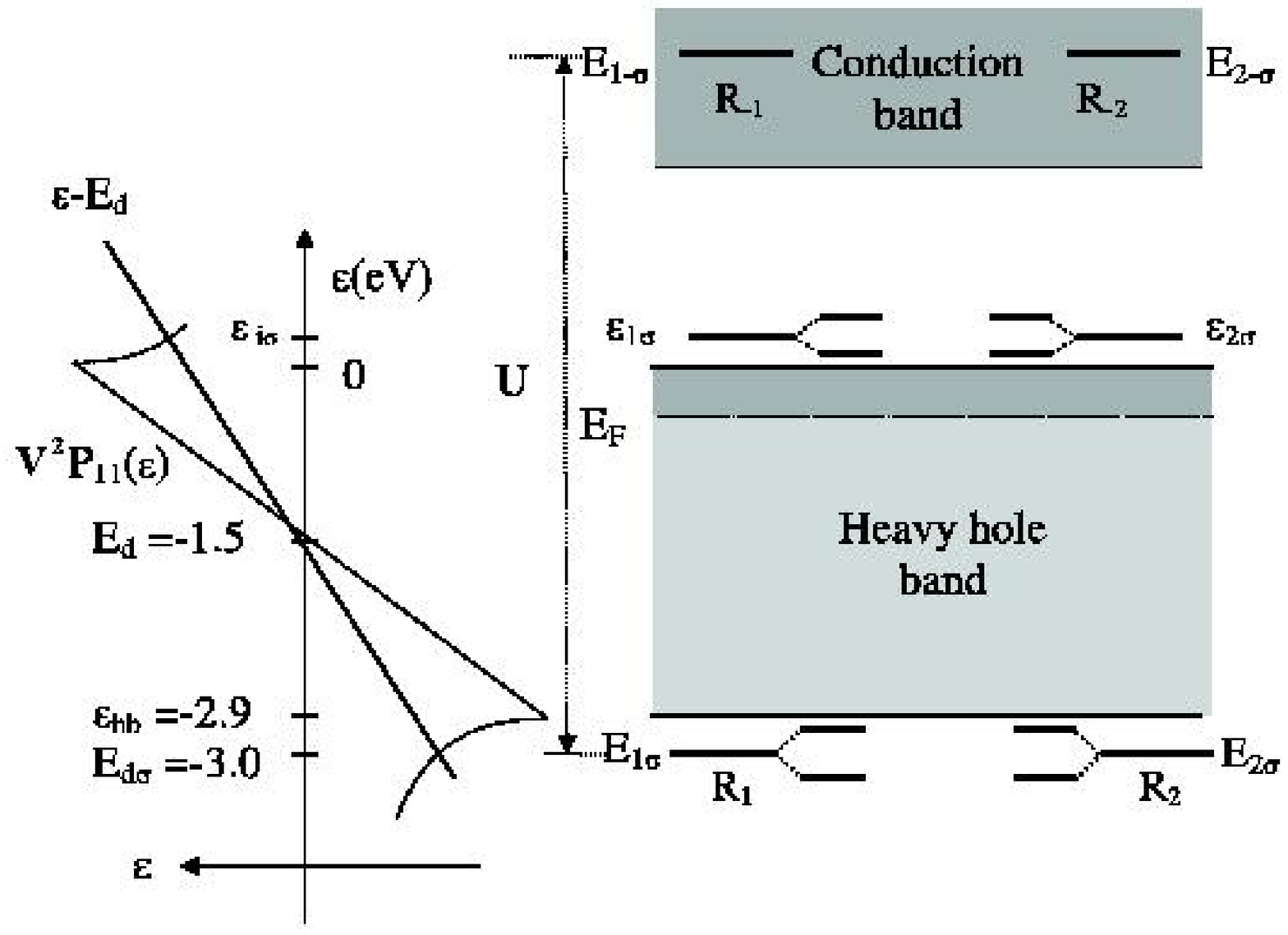}
\caption{Left panel: graphic solution of equation (\ref{AAA2}) for the
bonding CFR and antibonding DBH levels. Right panel:\ energy levels in
GaAs:Mn. The CFR\ d-levels $d^{5}/d^{4}$(denoted by R$_{1,2}$) of each
impurity, lie below the \textit{hh } band. The DBH\ levels (energies $%
\protect\varepsilon _{1\protect\sigma },$ $\protect\varepsilon _{2\protect%
\sigma }$) are split from the \textit{hh} band and form localized (acceptor)
levels in the energy gap. The CFR levels $d^{6}/d^{5}$ R$_{-1,-2}$ lie high
in the conduction band. }
\label{f.1}
\end{figure}

The zeros of the expression in the square brackets describe the impurity $d$%
-level renormalized by their hybridization with the \textit{hh} band. For
sufficiently large hybridization (or narrow valence band) the DBH states
arise above the top of the valence band, whereas the CFR\ levels appear deep
in the valence band below the bottom of the \textit{hh} subband (the left
panel of Fig.~\ref{f.1}). The inter-site interaction $~V^{4}|L_{12}^{\sigma
}|^{2}$ results in mixing of the CFR and the DBH states belonging to the two
impurities (right panel of Fig.~\ref{f.1}).

The occupied CFR levels $E_{i\sigma }$\ correspond to the states $d^{5}/d^{4}
$ of the Mn ions, whereas the empty $d^{6}/d^{5}$ CFR levels ($E_{i\overline{%
\sigma }}$) are shifted to the conduction band by the Anderson-Hubbard
repulsion $U$ (Fig.~\ref{f.1}) responsible for the spin-dependent
inter-impurity interaction and eventually for the FM order.

Both in GaAs:Mn and GaP:Mn the deep CFR states are completely occupied, and
the DBH states in the energy gap are empty (see, \textit{e.g.}, Refs. \cite
{Kikoin,Singh}). Two competing mechanisms of the magnetic interaction arise,
since the indirect exchange between neighboring impurities involves either
the empty states near the top of the valence band and the empty ($d^{6}/d^{5}
$) CFR levels. To determine the type of magnetic ordering, we consider the
indirect exchange between two neighboring magnetic ions and calculate its
sign and magnitude.

The impurity related correction to the energy of the system is given by the
standard formula
\begin{equation}
E^{magn}=\frac{1}{\pi }\mathrm{Im}\int_{-\infty }^{\infty }\varepsilon
\Delta \mathrm{Tr}\ \mathrm{G}[\varepsilon -i\delta \mathrm{sign}%
(\varepsilon -\varepsilon _{F})]d\varepsilon -\frac{1}{2}\sum_{i}U\bar{n}%
_{di\uparrow }{\bar{n}}_{di\downarrow },  \label{calc1}
\end{equation}
where $\mathrm{G}\left(E \right) =(E -H)^{-1}$ is the full Green's function.
We estimate first the contribution of the CFR levels. In this case the
two-site lattice Green's function $L_{12}(\varepsilon )\approx $ $%
L_{12}(E_{b\sigma })\equiv L$ is approximated by its value for the CFR $d$
-level position

\begin{equation}
E_{i\sigma }=E_{d}+V^{2}L_{ii}(E_{i\sigma }).  \label{deep}
\end{equation}
The quantity $L$ depends exponentially on the intersite distance $R_{12}$: $%
L\sim \exp (-\kappa _{b}R_{12})/\left( \kappa _{b}R_{12}w\right) $ where $%
\kappa _{b}= \sqrt{2m \left( \varepsilon _{b_{h}} - E_{i\sigma }\right) }/
\hbar $, with $\varepsilon _{bh}$ standing for the bottom of the \textit{hh}
band, $w$ is the \textit{hh }bandwidth, $E_{i\sigma }$ is the energy of the
CFR\ level below the bottom of the \textit{hh} band.

Due to the on-site repulsion \textit{U}, the structures of discrete CFR
levels differ for FM and antiferromagnetic (AFM) orientations of neighboring
impurity spins. For the AFM alignment the occupation numbers are $\bar{n}%
_{1\uparrow }=\bar{n}_{2\downarrow }=1,$ $\bar{n}_{1\downarrow } = \bar{n}%
_{2\uparrow }=0$ . As a result, the tunneling processes which influence
energy positions of the occupied states involve large $U$, and the secular
equation (\ref{AAA2}) yields the following expressions for the occupied
impurity levels (see Fig.~\ref{f.1}): \ $E_{b\uparrow }\approx E_{1\uparrow
}-J_{A},E_{b\downarrow }\approx E_{2\downarrow }-J_{A}$, where the indirect
exchange parameter
\begin{equation}
J_{A}=V^{4}L^{2}/U\;.  \label{iae}
\end{equation}
is the Anderson superexchange, which favors AFM order in transition metal
oxides.

In the FM case: $\bar{n}_{1\uparrow }=\bar{n}_{2\uparrow }=1,$ $\bar{n}%
_{1\downarrow }=\bar{n}_{2\downarrow }=0$. Then the states, which mediate
exchange between the impurity spins, are the mobile hole states below the
top of the valence band and the localized DBH states above its top. The role
of these states in the DMS is the similar to the that of the empty atomic
states in the conventional Zener double exchange mechanism for (La, A$^{2+}$%
)MnO$_{3}$ \cite{Zener}. An important difference is that in Zener's case the
Mn ions are in different valence states (Mn$^{3+}$ and Mn$^{4+}$). In other
words, one of the two levels $E_{b\uparrow }$ is empty. Since in our case
both of these states are occupied, the Zener mechanism in its original form
does not work. Formally, one gets a pair of bonding/antibonding CFR states $%
E_{(b,a)\uparrow }=E_{i\uparrow }\pm J_{Z}$ from (\ref{AAA2}) without any
energy gain, since both levels are occupied. It will be shown below that FM
order arises only at a finite hole concentration in the valence band.

Impurity related corrections to the band energy are obtained from the
integrations in Eq. (\ref{calc1}). (\textit{cf}. Ref. \cite{Haldane} where a
similar procedure was carried out for a single impurity). Then, the
variation of the total band energy due to the two-impurity scattering is
\begin{equation}
\Delta E_{b}=-\frac{1}{\pi }\mathrm{Im}\int_{\varepsilon _{hb}}^{\varepsilon
_{F}}d\varepsilon \ln \{[g_{d}^{-1}(i\varepsilon )-V^{2}L_{11}(i\varepsilon
)]^{2}-V^{4}L_{12}^{2}(i\varepsilon )\}.  \label{calc4}
\end{equation}
Here $L_{ij}(i\varepsilon )=P_{ij}(\varepsilon )+\frac{i}{2}\Gamma
_{ij}(\varepsilon )$ and
\begin{eqnarray}
P_{11}(\varepsilon ) &=&P_{22}(\varepsilon )=\int d\omega \frac{\rho (\omega
)}{\varepsilon -\omega },\;\;P_{12}(\varepsilon )=\int d\omega \frac{\sin
kR_{12}}{kR_{12}}\frac{\rho (\omega )}{\varepsilon -\omega },  \nonumber \\
\Gamma _{12}(\varepsilon ) &\approx &2\pi \rho (\varepsilon )\sin
kR_{12}/kR_{12},\ \ \ \Gamma _{11}(\varepsilon )\approx \pi \rho
(\varepsilon )=\Gamma _{22}(\varepsilon ).  \label{selfi}
\end{eqnarray}
Only the spin-up (majority spin) band  contributes to $\Delta
E_{b}$. Here and below the spin index is omitted for the sake of
brevity. The value of the wave-vector $k$ is found from the
equation $\varepsilon =\varepsilon _{hh}(k)$, where $\varepsilon
_{hh}(k)$ is the \textit{hh} energy dispersion. Since $V/\left(
g_{d}^{-1}-V^{2}P_{11}\right) =V/\left( \varepsilon
-E_{d}-V^{2}P_{11}\left( {\varepsilon }\right) \right) \ll 1$ one
can extract from Eq. (\ref{calc4}) the spin-dependent \textit{hh}
band contribution to the exchange energy, which reads

\begin{equation}
E_{ex}=-\frac{V^{4}}{4\pi }(\overrightarrow{\mathbf{\sigma }}_{1}\cdot
\overrightarrow{\mathbf{\sigma }}_{2}+3)\left\{ \int_{\varepsilon
_{F}}^{0}d\varepsilon \frac{\Gamma _{12}(\varepsilon )P_{12}(\varepsilon )}{
[\varepsilon -E_{d}-V^{2}P_{11}({\varepsilon })]^{2}+\frac{V^{4}}{4}\Gamma
_{11}^{2}}+3x\;\frac{P_{12}(\varepsilon _{i})P_{12}^{\prime }(\varepsilon
_{i})}{\left[ 1-V^{2}P_{11}^{\prime }(\varepsilon _{i})\right] ^{2}}\right\}
\label{Ehol}
\end{equation}
where $\overrightarrow{\mathbf{\sigma }}_{1,2}$ are the vectors of Pauli
matrices. The terms in curly brackets correspond to the contribution of
mobile and localized holes. The factor 3 appeared due to the degeneracy of
the localized acceptor $p$-levels $\varepsilon _{i}.$ The function $%
P_{ij}^{^{\prime }}(\varepsilon )=dP_{ij}^{{}}(\varepsilon )/d\varepsilon $
is negative at $\varepsilon =\varepsilon _{i}$; (see left panel of Fig.~\ref
{f.1}).

Eq. (\ref{Ehol}) is the main result of our theory from which we obtain $T_{C}
$. In the evaluations of Anderson-type and Zener-type coupling constants we
use the estimates: $P_{ij}\sim w^{-1},\;\Gamma _{ij}\sim \varepsilon
_{F}^{1/2}w^{-3/2}$, $\varepsilon _{F}-E_{d}-V^{2}P_{11}({\varepsilon }%
_{F})=4\alpha V^{2}/w$ with $\alpha <1$ (see left panel of Fig.~ \ref{f.1}).
Then one finds from Eqs. (\ref{iae}) and (\ref{Ehol}) that $J_{A}\;\sim V^{4}%
\left[ \exp \left( -\kappa _{b}R_{12}\right) /\left( \kappa
_{b}R_{12}\right) \right] ^{2}/(Uw^{2})$, with $\kappa _{b}=\sqrt{2m\left(
\varepsilon _{b_{h}}-E_{d\sigma }\right) }/\hbar $, $J_{F}\;\sim
2\varepsilon _{F}\left( \varepsilon _{F}/w\right) ^{1/2}/\left[ \left(
4\alpha \right) ^{2}+\varepsilon _{F}/\left( 4w\right) \right] $ and FM
pairing is realized provided $J_{F}>J_{A}.$

\begin{figure}[th]
\includegraphics[width=14cm]{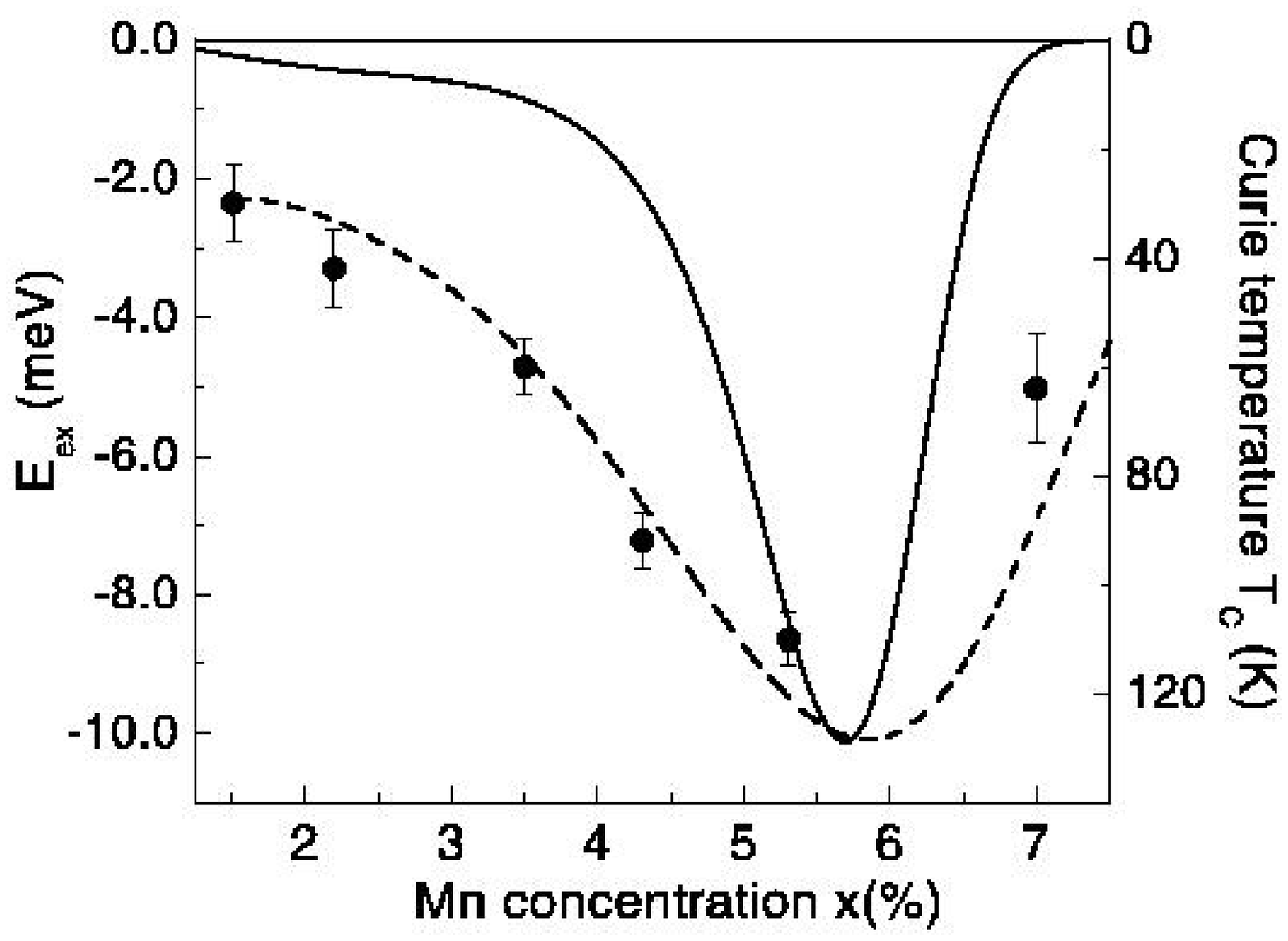}
\caption{The dependence of the kinematic exchange (left axis) and the $T_{C}$
(right axis) on the impurity concentration. Experimental results (filled
circles with error bars) are taken from Ref.\emph{\ }\protect\cite
{fmatsukara}. The solid and dashed curve are obtained using the experimental
obtained central and upper values of the error bars for the hole density,
respectively of Ref. \protect\cite{fmatsukara}}
\label{f.2}
\end{figure}

To calculate the exchange energy Eq.(\ref{Ehol}), one needs the dependence
of the Fermi level on the Mn concentration $\varepsilon _{F}(x)$. It is
governed by the equation $x_{s}=2\int_{\varepsilon _{F}}^{0}\rho
_{v}(\varepsilon )d\varepsilon ,$ where $x_{s}$ is the hole concentration
per site. We assume a semi-elliptical density of states, $\rho
_{v}(\varepsilon )=2\left[ \pi w^{2}\right] ^{-1}\sqrt{-\varepsilon
(\varepsilon +w)\theta (-\varepsilon )},$ where $w=2.9$ eV \cite{Mod}. The
hole concentration per site, $x_{s}$, is proportional to the hole density
per volume $p_{h}$: $x_{s}=1/8a^{3}p_{h}$, as there are four III-V pairs in
a unit cell volume $a^{3}$ in zinc-blende structures. Based on the
experimental data $p_{h}(x)$ of Ref.\cite{fmatsukara} we used a polynomial
fit for the hole density dependence on the Mn concentration in GaAs:Mn.

To compare the values of $J_{A}$ and $J_{F}$ we use the calculated $%
\varepsilon _{F}(x=5.3\%)=-50$meV, $U\approx 4.5$ eV, the hybridization
parameter $V=1.27$eV obtained from (\ref{deep}) for the acceptor level $%
\varepsilon _{i}=85$ meV ( $\varepsilon _{i}^{\exp }=110$ meV \cite{Zunger})
and the CFR\ level $E_{i}=-3.0$eV ($E_{i}^{\exp }=-3.4$eV \cite{Asklund}).
At these values $\alpha =w\left[ \varepsilon _{F} - E_{d} - V^{2} P_{11}( {%
\varepsilon }_{F})\right] /\left( 4V^{2}\right) =-0.32$ and the ratio $r=2.13
$ justifies the dominance of FM coupling in GaAs:Mn: $J_{F}>J_{A}$

\begin{figure}[th]
\includegraphics[width=14cm]{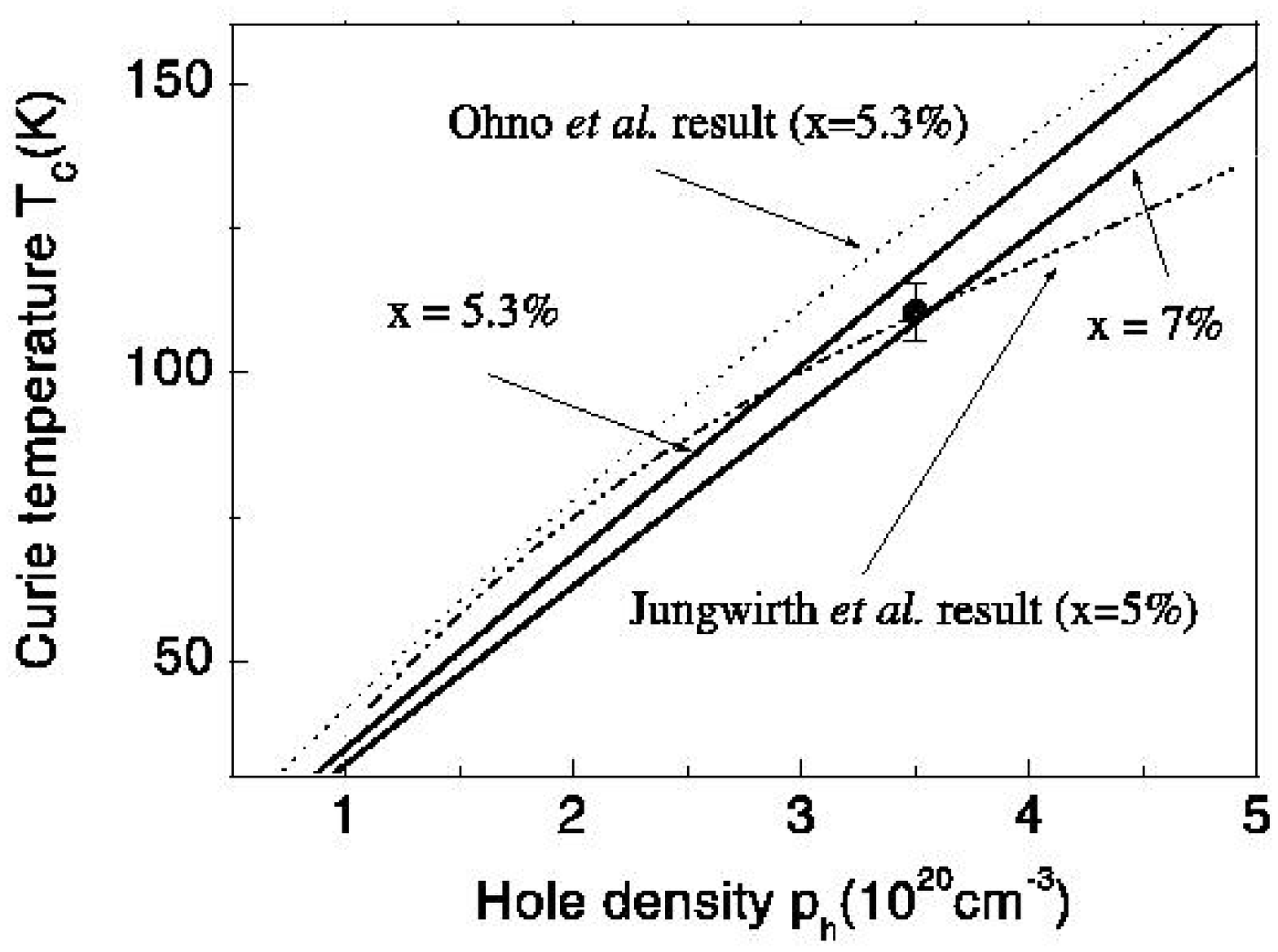}
\caption{The predicted dependence $T_{C}$\ $\left( p_{h}\right) $, at the
fixed manganese concentration $x=5.3\%$ and 7\%. (solid lines). The dashed
line are the theoretical result of Ref.~ \protect\cite{Ohno2}, derived from
the phenomenological Landau free-energy functional within the framework of
the Zener model. The dash-dot line is the recent result \protect\cite{Jung}
obtained by using the exchange-enhanced mean-field theory. The point at $%
x=5.3\%$ corresponds to the experimentally found \protect\cite{fmatsukara}
maximal value of $T_{C}$ for GaAs:Mn.}
\label{f.3}
\end{figure}

Eq. (\ref{Ehol}) allows one to compute the Curie temperature $%
T_{C}(x)=E_{ex}(x)/k_{B}$. The results are presented in Fig.~\ref{f.2}. The
calculated dependence $T_{C}(p_{h})$ is given in Fig.~\ref{f.3} for
different values of the Mn concentration.

Although we neglected in this paper formation of an impurity band around $%
x\approx 5\%$ our results for $T_{C}(x)$ and $T_{C}(p_{h})$ are in good
agreement both with available experimental data, and the theoretical
phenomenological estimates. The dielectric structure parameters such as DBH,
lower CFR energies and $V$ are calculated self-consistently (Eq. (\ref{deep}%
) and see the left panel of Fig.\ref{f.1}) for the graphical solution). Even
our estimated value of $V=1.27~eV$ is in a reasonable agreement with the
hybridization parameter $\sim 1.4~eV$ which can be extracted from the data
of Ref.\cite{Okabayashi}. This implies that the present theory, in fact, has
no real fitting parameters. Note that in Ref. \cite{Ohno2} the authors took
into account the magnetic correlation mediated by holes originating from
shallow acceptors without a proper regard of the role of the impurity
levels, which we consider to be important, and later calculated $T_{C}$ by
minimizing the Landau free energy.

In conclusion, we proposed a microscopic model for a double exchange in
\textit{p}-type III-V:Mn DMS based on the known mechanism of interaction
between substitutional transition metal impurities and the host
semiconductor \cite{Zunger}, \cite{Kikoin}. A presence of holes is crucial
for the FM double exchange between neighboring Mn ions. Our model there does
not require an introduction a \emph{adhoc} phenomenological \textit{pd}- or
RKKY-type exchanges. The source of the magnetic coupling is the energy gain
in the kinetic energy of holes in case when neighboring ion spins are
parallel. Among existing approaches the closest one to ours is the LSDA+U
method\cite{sanvito}, in which the system of $t_{2}\sigma $- and $e\sigma $-
levels and the hole pocket for majority spin density of states correlates
with our spectrum presented in Fig.~\ref{f.1}. However, instead of
extracting the \textit{pd}-exchange from this band structure we calculated
the genuine Zener-like exchange which emerges since the band energy is lower
in the FM case then in the AFM case \cite{akai}. In our model this exchange
is defined by Eq. (\ref{Ehol}). Similar ideas were applied to DMS CdGeP$_{2}$%
:Mn (\cite{maha}), where the interplay between CFR and DBH states turned out
to be crucial for the onset of FM order. The theory may be applied to GaP:Mn
which has a similar structure of chemical bonds around the Mn ion \cite
{Singh}; however, experimental data on the CFR and DBH states as well as the
dependencies $p_{h}\left( x\right) $ and $T_{C}(x)$ are not available. The
proposed \textit{kinematic exchange} can also be applied to other DMS
including GaN:Mn and A$^{2+}$GeP$_{2}$:Mn, which is left for future work.

\acknowledgments This work was supported by the Flemish Science Foundation
(FWO-VI), the Belgian Inter-University Attraction Poles (IUAP), the
``Onderzoeksraad van de Universiteit Antwerpen'' (GOA) and the Israel
Physical Society.


\begin{thebibliography}{99}

\bibitem{Ohno10} H. Ohno, A. Shen, F. Matsukara, A. Oiwa, A. Endo,
S. Katsumoto, Y. Iye, Appl. Phys. Lett. {\bf 69}, 363, (1996).

\bibitem{Reed} M.L. Reed, N.A. El-Masry, H.H. Stadelmaier, M.K.
Ritums, M.J. Reed, C.A. Parker, J.C. Roberts, S.M. Bedair, Appl.
Phys. Lett., {\bf 79}, 3473 (2001); N. Theodoropoulou, A.F.
Hebard, M.E. Overberg, C.R. Abernathy, S.J. Pearton, S.N.G. Chu,
R.G. Wilson, cond-mat/0201492 (2002).

\bibitem{Dietl}  Yu.G. Semenov and S.M. Ryabchenko, Low Temp.
Phys. [Fizika nizkih temperatura - {\em in Russian}], {\bf 26},
1197 (2000); T. Dietl, H. Ohno and F. Matsukara, Phys. Rev. B,
{\bf 63}, 2034 {\bf 2001}; J. Inoue, S. Nonoyama and H. Itoh,
Phys. Rev. Lett., {\bf 85}, 4610 (2000).

\bibitem{Duga} V.K. Dugaev, V.I. Litvinov, J. Barna\'{s}, and M.
Vieira, cond-mat/0203119 (2002).

\bibitem{Zener} C. Zener, Phys. Rev., {\bf 82}, 403 (1951).

\bibitem{Anderson} P.W. Anderson, Phys. Rev., {\bf 124}, 41 (1961).

\bibitem{fleurov1} V.N. Fleurov and K.A. Kikoin, J. Phys. C:
Solid State Phys. Phys., {\bf 9}, 1673 (1976).

\bibitem{Haldane} F.D.M. Haldane and P.W. Anderson, Phys.
Rev. B {\bf 13}, 2553 (1976).

\bibitem{Alex} S. Alexander and P.W. Anderson, Phys. Rev. A, {\bf
133}, 1594 (1964).

\bibitem{Zunger} A.Zunger, Solid State Physics., Eds. H.
Ehrenreich and D. Turnbull, Vol {\bf 39} (Academic Press, Orlando)
(1986), p. 276.

\bibitem{Kikoin}  K.A. Kikoin and V.N. Fleurov, Transition
Metal Impurities in Semiconductors, (World Scientific Publishing,
Singapore) (1994).

\bibitem{Hemstreet} L.A. Hemstreet, Phys.Rev. B {\bf 22}, 4590 (1980).

\bibitem{Picoli} G. Picoli, A. Chomette, and M. Lannoo,
Phys.Rev. B {\bf 30}, 7138 (1984).

\bibitem{Singh} V.A. Singh and A. Zunger, Phys. Rev. B {\bf 31},
3729 (1985).

\bibitem{omiya} D.K. Omiya, A. Matsukara, Y. Shem, Y. Ohno and H.
Ohno, Physica E {\bf 10}, 206 (2000).

\bibitem{fmatsukara}  F. Matsukara, H. Ohno, and A. Shen, Phys.
Rev. B {\bf 57}, R2037 (1998).

\bibitem{Mod} O. Madelung, Data in Science and Technology -
Semiconductors: Group IV Elements and III-V Compounds. Ed.: R.
Poerschke, (Springer Verlag, Berlin) (1991) p. 101.

\bibitem{Asklund} H. Asklund, L. Ilver, and J. Kanski,
cond-mat/0112287 (2001).

\bibitem{Okabayashi} J. Okabayashi, A. Kimura, O. Rader,
Y.T. Mizokawa, A. Fujimori, T. Hayashi, M. Tanaka, Phys. Rev. B
{\bf 58}, R4211 (1998).

\bibitem{Ohno2} H. Ohno and F. Matsukara, Solid
State Commun. {\bf 117}, 179 (2001).

\bibitem{sanvito} S. Sanvito, P. Ordejon, and N.A. Hill, Phys.
Rev. B {\bf 63}, 165206 (2001).

\bibitem{akai} Akai H., Phys. Rev. Lett., {\bf 81}, 3002 (1998).

\bibitem{maha} P. Mahadevan P. and A. Zunger, Phys.Rev.Lett., {\bf
88}, 047205 (2002).

\bibitem{Jung} T. Jungwirth, J. K\"{o}nig, J. Sinova,  J. Kucera, A.H.
MacDonald, cond-mat/0201157 (2002).
\end{thebibliography}
\end{document}